\documentstyle[12pt]{article}

\def\beq{\begin{equation}}
\def\eeq{\end{equation}}
\def\barr{\begin{eqnarray}}
\def\earr{\end{eqnarray}}
\def\cat{{\cal T}}
\def\cap{{\cal P}}
\def\catp{{\cal T}'}
\def\capp{{\cal P}'}
\def\app{A_{\pi \pi}}
\def\apk{A_{\pi K}}
\def\abpp{\overline{A}_{\pi \pi}}
\def\abpk{\overline{A}_{\pi K}}

\def\tcapp{\tilde{\cal P}'}
\def\tru{\tilde{r_u}}
\def \ite{{\it et al.}}

\def \cp89{{\it CP Violation,} edited by C. Jarlskog (World Scientific,
Singapore, 1989)}

\def \f79{{\it Proceedings of the 1979 International Symposium on Lepton and
Photon Interactions at High Energies,} Fermilab, August 23-29, 1979, ed. by
T. B. W. Kirk and H. D. I. Abarbanel (Fermi National Accelerator Laboratory,
Batavia, IL, 1979}
\def \hb87{{\it Proceeding of the 1987 International Symposium on Lepton and
Photon Interactions at High Energies,} Hamburg, 1987, ed. by W. Bartel
and R. R\"uckl (Nucl. Phys. B, Proc. Suppl., vol. 3) (North-Holland,
Amsterdam, 1988)}

\def \ibj#1#2#3{~{\bf#1}, #2 (#3)}
\def \ichep72{{\it Proceedings of the XVI International Conference on High
Energy Physics}, Chicago and Batavia, Illinois, Sept. 6 -- 13, 1972,
edited by J. D. Jackson, A. Roberts, and R. Donaldson (Fermilab, Batavia,
IL, 1972)}

\def \lkl87{{\it Selected Topics in Electroweak Interactions} (Proceedings of
the Second Lake Louise Institute on New Frontiers in Particle Physics, 15 --
21 February, 1987), edited by J. M. Cameron \ite~(World Scientific, Singapore,
1987)}

\def \np#1#2#3{Nucl. Phys. {\bf#1}, #2 (#3)}

\def \plb#1#2#3{Phys. Lett. B {\bf#1}, #2 (#3)}

\def \prd#1#2#3{Phys. Rev. D {\bf#1}, #2 (#3)}
\def \prl#1#2#3{Phys. Rev. Lett. {\bf#1}, #2 (#3)}

\def \si90{25th International Conference on High Energy Physics, Singapore,
Aug. 2-8, 1990}
\def \slc87{{\it Proceedings of the Salt Lake City Meeting} (Division of
Particles and Fields, American Physical Society, Salt Lake City, Utah, 1987),
ed. by C. DeTar and J. S. Ball (World Scientific, Singapore, 1987)}

\def \slac89{{\it Proceedings of the XIVth International Symposium on
Lepton and Photon Interactions,} Stanford, California, 1989, edited by M.
Riordan (World Scientific, Singapore, 1990)}
\def \smass82{{\it Proceedings of the 1982 DPF Summer Study on Elementary
Particle Physics and Future Facilities}, Snowmass, Colorado, edited by R.
Donaldson, R. Gustafson, and F. Paige (World Scientific, Singapore, 1982)}
\def \smass90{{\it Research Directions for the Decade} (Proceedings of the
1990 Summer Study on High Energy Physics, June 25--July 13, Snowmass, Colorado)\
,
edited by E. L. Berger (World Scientific, Singapore, 1992)}
\def \stone{{\it $B$ Decays} (Revised 2nd Edition), edited by S. Stone,
World Scientific, Singapore, 1994)}
\def \tasi90{{\it Testing the Standard Model} (Proceedings of the 1990
Theoretical Advanced Study Institute in Elementary Particle Physics, Boulder,
Colorado, 3--27 June, 1990), edited by M. Cveti\v{c} and P. Langacker
(World Scientific, Singapore, 1991)}

\def \zpc#1#2#3{Zeit. Phys. C {\bf#1}, #2 (#3)}

\begin{document}

\rightline{IC/98/210}
\rightline{hep-ph/9811506}

\bigskip

\begin{center}

{\Large \bf Flavor SU(3) in Hadronic B Decays} 
\footnote{Talk given at The Workshop
on B and Neutrino Physics, Mehta Research Institute, Allahabad,
1998.}

\bigskip

{\it Amol Dighe \\ The Abdus Salam International Center for
Theoretical Physics \\ 34100, Trieste, Italy.}

\end{center}

\bigskip






\begin{abstract}
	Here we shall outline a few methods that use the flavor SU(3)
symmetry in the decays of $B$ mesons 
to determine the angles of the unitarity triangle
and to identify the decay modes which
would display a significant CP violation.
\end{abstract}

\section{Introduction}

	The decays of $B$ mesons are crucial in answering the 
question of whether the CKM matrix \cite{ckm}
can describe all the CP violation that we observe or that will be
observed. This is an important question, since it probes the very
basis of the charged current weak interactions and the origin 
of CP violation. The many decay modes available in $B$ decays 
provide consistency checks, which in addition probe the standard
model from various angles. Thus, possibilities for the first and the 
surest signals of the physics beyond the standard model lie in
the analyses of these decay modes.

	The determination of the angles of the unitarity triangle
is one of the goals of the theoretical and experimental
efforts being put in this field. The search for theoretically
clean and experimentally feasible decay modes is still on. The 
flavor SU(3) symmetry \cite{zepp, savagesu3, chau}, 
which should approximately hold in 
$B$ decays, shows us some paths in this search.

\section{$B \to PP$ amplitude polygons without time information}
\label{rigid}

	Let us consider the decays of a $B$ meson into two light 
pseudoscalars, $P_1$ and $P_2$.
	Without time measurements, the only available information
from a decay mode is its total decay rate, and hence the magnitude
of its amplitude. In order to be able to
measure the relative phases between amplitudes, we then need some
theoretical relations in the form of triangles (in the
complex plane) whose sides will
be the amplitudes of these decay rates. The angles of 
triangles are determined given the lengths of their sides, 
and thus the relative
phases between the amplitudes are known given their magnitudes. 
The next step is going
from these relative phases to the angles of the unitarity 
triangle, $\alpha, \beta$ and $\gamma$ (also called as the 
CKM phases).

	Two approaches have been taken to get the theoretical
amplitude triangle (or quadrangle) relations. For the
decays with $B$ going into two light pseudoscalars, the total
amplitude may be expressed either in the basis of six 
``Feynman diagrams'' $T$ (tree), $C$ (color-suppressed tree),
$P$ (penguin), $E$ (exchange), $A$ (annihilation) and $EA$
(penguin annihilation) \cite{glr,ghlrd50}; or in the basis of 
six SU(3) invariant amplitudes \cite{savagesu3,desh-he}. 
Both these approaches are equivalent. (Actually, only
five of these six
amplitudes are independent, as shown in \cite{zepp}.)
After neglecting the
annihilation-type contributions
(which are expected to be suppressed by a factor of $f_B/m_B 
\approx 5 \%$), numerous equivalence, triangle and quadrangle
relations are obtained \cite{zepp,glr,ghlrd50,desh-he,triangles,
hierarchy,art,etaetap}. The decay modes need
to be divided into two groups, $\Delta S = 0$ and $|\Delta S|=1$.
In a triangle, only modes from within the same group 
should appear, since
the dominant (tree and penguin) amplitudes in these two groups
contribute with different CKM strengths. Within each of these
groups, all the decay modes of the type $(B^+ / B^0 / B_s)
\to P_1 P_2$ form the sides of at least one of the connected
triangles, giving rise to ``rigid polygons'' \cite{art}. 
Even in the presence of singlet-octet mixing in the
$\eta-\eta'$ system, it is still possible to extend the
formalism and obtain amplitude triangle relations 
\cite{art,etaetap}.

	The process of going from the amplitude triangles to the 
CKM phases is not always theoretically clean and estimations of
the relative strengths of various contributions are needed in order
to decide whether some of them can be neglected to simplify
the relations. The hierarchy of amplitudes \cite{hierarchy}
is different for $\Delta S= 0$ and $|\Delta S| = 1$ modes due to 
the different CKM factors.

	The amplitudes for these two types of modes may be
written approximately as
\barr
A(\Delta S = 0) & =  V_{tb}^* V_{td} P + V_{ub}^* V_{ud} T
		& =  e^{-i\beta} e^{i\delta_P} \cap +
			 e^{i\gamma} e^{i\delta_T} \cat \\
A(\Delta S = 1) & =  V_{tb}^* V_{ts} P' + V_{ub}^* V_{us} T' 
		& =  e^{i\pi} e^{i\delta_P} \capp +
			 e^{i\gamma} e^{i\delta_T} \catp  ~~.
\earr
where primed quantities denote $|\Delta S| = 1$ decays, $\cap,
\cat, \capp, \catp$ are the magnitudes of the respective 
contributions, $\delta_P,~ \delta_T$
are strong phases and $\beta, \gamma, \pi$ are the weak phases
from the CKM matrix elements. The corrections
due to quarks other than the top quark to the penguin
amplitudes have been neglected here. 
These corrections have been
estimated in \cite{charmedP}.
$T$ and $P$ here may contain some additional contributions due to 
electroweak penguins, but that does not affect the construction
of triangles \cite{art,etaetap}. 

For the tree diagram, the first order SU(3) corrections may be 
introduced through
the $K/\pi$ ratio of decay constants
\beq
\label{t-fact}
{\cal T'}/ {\cal T}= |V_{us}/V_{ud}| (f_K / f_{\pi})\equiv
r_u(f_K / f_{\pi}) \equiv \tilde{r}_u~~.
\eeq
Here factorization has been used 
for the tree contribution, which is supported by experiments
\cite{Browder,BS} and
justified for $B \to \pi \pi$ and $\pi K$ by the high momentum with
which the
two color-singlet mesons separate from one another.
Since factorization is
questionable for
penguin amplitudes, it has not been used, but it is assumed 
that the phase $\delta_P$ is unaffected by the SU(3)
breaking. Since 
this phase is likely  to be small \cite{Penphase}, this assumption is
not expected to introduce a significant uncertaintly in the determination
of the weak phases.

The $P'$ term is expected to dominate the decays of the type 
$|\Delta S| = 1$ and $T$ term would dominate the $\Delta S = 0$
decays \cite{hierarchy}. 
Separate strategies can then be proposed for obtaining
the phases from the two types of decays, which I shall illustrate
with an example from each case. The details can be found in 
\cite{art}.

\subsection{$\Delta S = 0$ :}

	Once the triangles have been constructed, rotate them
such that the amplitudes with only penguin contributions lie
along the x-axis. The (rotated) amplitude of a decay mode
with nonzero tree contribution will then be 
\beq
A_R(\Delta S = 0) = {\cal P} + e^{i (\beta+\gamma)} 
	e^{i(\delta_T-\delta_P)} {\cal T}~~.
\eeq
When the corresponding antiparticle triangle is similarly
constructed and aligned, the amplitude of the corresponding
CP-conjugate decay will be
\beq
\overline{A}_R(\Delta S = 0) = {\cal P} + e^{-i (\beta+\gamma)} 
	e^{i(\delta_T-\delta_P)} {\cal T}~~.
\eeq
If the penguin contribution is much smaller than
the tree one, the angle between $A_R$ and $\overline{A}_R$
is $2(\beta + \gamma) = 2\pi - 2\alpha \equiv -2\alpha$.

\subsection{$|\Delta S| = 1$ :}

With the same reorientation as above (amplitudes with only penguin
contributions along the x axis), we have
\beq
A_R(\Delta S = 1) = - {\cal P'} + e^{i\gamma} 
	e^{i(\delta_T-\delta_P)} {\cal T'}~~
\eeq
and
\beq
\overline{A}_R(\Delta S = -1) = - {\cal P'} + e^{-i\gamma} 
	e^{i(\delta_T-\delta_P)} {\cal T'}~~.
\eeq
Then
\beq
A_R(\Delta S = 1) - \overline{A}_R(\Delta S = -1)=
	2 i \sin \gamma ~ e^{i(\delta_T-\delta_P)} {\cal T'}~~.
\eeq
Using Eq.~(\ref{t-fact}) and the value of ${\cal T}$ obtained
from a tree-dominated $\Delta S= 0$ decay, we can obtain
$\gamma$ as well as $ \delta_T-\delta_P$.

\section{Decays to kaons and charged pions with time information}

	If the measurements of time-dependent rates for
$B^0(t) \to \pi^+ \pi^-$ and  $\overline{B}^0(t) \to \pi^+ \pi^-$ 
are added to the measurements of the total rates for 
$B^0 \to \pi^- K^+$, $\overline{B}^0 \to \pi^+ K^-$ and 
$B^{\pm} \to K_S \pi^{\pm}$, the CKM phases may be determined
\cite{gr, dgr} without having to detect $\pi^0$ as follows:

Neglecting color-suppressed electroweak penguin effects of order
$|P'^C_{EW}/P'|={\cal O}((1/5)^2)$,
we can write
$$
\app \equiv A(B^0 \to \pi^+ \pi^-) = 
{\cal T}e^{i\delta_T}e^{i\gamma}+{\cal P}e^{i\delta_P}e^{-i\beta}~~,
$$
$$
\apk \equiv A(B^0 \to \pi^- K^+) =
\tilde{r}_u{\cal T}e^{i\delta_T}e^{i\gamma}-\tilde{\cal P'}
e^{i\delta_P}~~,
$$
\beq 
A_+ \equiv A(B^+ \to \pi^+ K^0) = \tilde{\cal P'}e^{i\delta_P}~~.
\label{eqn:withbk}
\eeq

The magnitude of the $\Delta S = 1$ penguin amplitude has been
denoted by
$\tcapp$, to allow for SU(3) breaking. The numerous 
{\it a priori} unknown parameters in
(\ref{eqn:withbk}), including the two weak phases
$\gamma$ and $\alpha\equiv\pi-\beta-\gamma$, can be determined from
the rate
measurements of the above three processes and their
charge-conjugates.

The amplitudes for the corresponding charge-conjugate decay processes
are
simply obtained by changing the signs of the weak phases $\gamma$ and
$\beta$.
We denote the charge-conjugate amplitudes corresponding to
(\ref{eqn:withbk})
by $\overline{A}_{\pi\pi},~\overline{A}_{\pi K},~A_-$, respectively.
A state
initially tagged as a $B^0$ or $\overline{B}^0$ will be called
$B^0(t)$ or
$\overline{B}^0(t)$.  The time-dependent decay rates of these states
to
$\pi^+\pi^-$ are given by
$$
\Gamma(B^0(t)\to\pi^+\pi^-)=e^{-\Gamma t}[|\app|^2\cos^2({\Delta
m\over
2}t)+|\abpp|^2\sin^2({\Delta m\over 2}t)
$$
$$
+{\rm Im}(e^{2i\beta} \app \abpp^*)\sin(\Delta mt)]~~,
$$
$$
\Gamma(\overline{B}^0(t)\to\pi^+\pi^-) = e^{-\Gamma
t}[|\app|^2\sin^2({\Delta
m\over 2}t) + |\abpp|^2\cos^2({\Delta m\over 2}t)
$$
\beq
-{\rm Im}(e^{2i\beta} \app \abpp^* \sin(\Delta mt)]~~.
\eeq

Measurements of these decay rates determine the quantities
$|\app|^2,~|\abpp|^2$ and
${\rm
Im}(e^{2i\beta} \app \abpp^*)$.  It is convenient to define sums and
differences of the first two quantities, and we find
$$
A \equiv \frac{1}{2}(|\app|^2 + |\abpp|^2)
= \cat^2 + \cap^2 - 2 \cat \cap \cos \delta \cos \alpha~~~,
$$
$$
B \equiv \frac{1}{2}(|\app|^2 - |\abpp|^2)
= - 2 \cat \cap \sin \delta \sin \alpha~~~,
$$
\beq \label{eqn:ABC}
C \equiv {\rm Im}~(e^{2 i \beta} \app \abpp^*) = - \cat^2 \sin
2\alpha
+ 2 \cat \cap \cos \delta \sin \alpha~~~,
\eeq
where we use $\beta + \gamma = \pi-\alpha$ and where we define
$\delta \equiv
\delta_T - \delta_P$.

The rates of the self-tagging modes $\pi^- K^+,~\pi^+ K^-$ and
$\pi^+K^0$ or
$\pi^- \overline{K}^0$ determine $|\apk|^2,~|\abpk|^2$ and $|A_+|^2$,
respectively.  Again, we can take sums and differences of the first two,
and find
$$
D \equiv \frac{1}{2}(|\apk|^2 + |\abpk|^2) = (\tru \cat)^2 + \tcapp^2
- 2 \tru
\cat \tcapp \cos \delta \cos \gamma~~~~,
$$
$$
E \equiv \frac{1}{2}(|\apk|^2 - |\abpk|^2) = 2 \tru \cat \tcapp \sin
\delta
\sin \gamma~~~,
$$
\beq \label{eqn:DEF}
F \equiv |A_+|^2 = |A_-|^2 = \tcapp^2~~~.
\eeq
The rates for $B^+ \to \pi^+ K^0$ and $B^- \to \pi^- \overline{K}^0$
are
expected to be equal, since only penguin amplitudes are expected to
contribute to these processes.  
Measurement of the six quantitities $A - F$ suffices to determine all
six
parameters $\alpha,~\gamma,~\cat,~\cap,~ \tcapp,~\delta$ up to
discrete
ambiguities. The accuracy to which they can be determined is
estimated in \cite{dgr} and the discrete ambiguities are 
studied in \cite{discrete}.
The CKM parameter $r_t\equiv|V_{ts}/V_{td}|$, which is
still
largely unknown, is obtained from the unitarity triangle in terms of
$\alpha$
and $\gamma$ as $r_u r_t= \sin\alpha / \sin\gamma$.

$B$ and $E$ are proportional to $\sin \delta$, and thus
would vanish in the absence of a strong phase difference.  In that
case, the number of equations becomes less than the number
of unknowns and one
would have to assume a relation between $\tilde{\cal P'}$ and ${\cal
P}$ or some
other constraint in order to obtain a solution.  

This problem may be overcome if the SU(3)-related decays
$B_s^0 \to \pi^+ K^-$, $B_s^0(t) \to K^+ K^-$ and 
$B_s^0 \to K^0 {\bar K^0}$are included \cite{london}.
One can then also get rid of some of the 
discrete ambiguities and $\gamma$ can be
obtained cleanly with no penguin contamination.
But a large number of quantities (12 decay rates)
need to be measured (some of them involving $B_s$), 
so it will be difficult to implement
this method in the near future.

\section{Angular distributions of $B$ decays to two vector mesons}

The decay modes of $B$ into two vector mesons, each of
which decay into two particles each, are very promising, mainly
because of the larger number of observables at one's disposal 
through the angular distributions 
\cite{angular, ddlr}
of the decays.
The time-dependent angular distributions contain information about 
the lifetimes, mass differences, strong and weak phases, 
form factors,
and CP violating quantities.

Let me illustrate with the example of 
$B_s \to J/\psi \phi$. With the angles $\theta, \phi, \psi$
defined as in \cite{ddf1}, the angular distribution is

$$
\frac{d^3 \Gamma [B_s(t) \to J/\psi (\to l^+ l^-) \phi (\to K^+
K^-)]}
{d \cos \theta~d \varphi~d \cos \psi}
\propto \frac{9}{32 \pi} \Bigl[~2 |A_0(t)|^2 \cos^2 \psi (1 - \sin^2
\theta
\cos^2 \varphi)
\hfill{ }
$$
$$
+ \sin^2 \psi \{ |A_\parallel(t)|^2  (1 - \sin^2 \theta \sin^2
\varphi)
+ |A_\perp(t)|^2 \sin^2 \theta - \mbox{ Im }(A_\parallel^*(t)
A_\perp(t))
\sin 2 \theta \sin \varphi \}~~~
$$
\beq \label{triple}
+\frac{1}{\sqrt{2}}\sin2\psi\{{\mbox{ Re }}(A_0^*(t) A_\parallel(t))
\sin^2
\theta
\sin2\varphi+\mbox{ Im }(A_0^*(t)A_\perp(t))\sin2\theta\cos\varphi~\}
~\Bigr]~~.
\eeq

The time evolutions of the coefficients of the angular terms is given
in Table~\ref{tab1}. Here $\Gamma_L$ and $\Gamma_H$ are the widths of
the light and heavy $B_s$ mass eigenstates, $B_s^L$ and $B_s^H$
respectively, and $\Delta m$ is the mass difference between them.
$\overline{\Gamma}$ is the average of $\Gamma_L$ and $\Gamma_H$.  Here
$\delta_1 \equiv {\rm Arg}(A_{\|}^{\ast}(0) A_{\perp}(0)) $ and $\delta_2
\equiv {\rm Arg} (A_0^{\ast}(0) A_{\perp}(0)) $ are the strong phases, and
$\delta \phi \approx 2 \lambda^2 \eta$ is related to an angle of a
(squashed) unitarity triangle ~\cite{dphi}, which is very small in the
standard model [ $\approx (0.03)$].

\begin{table}[t]
\begin{center}
\begin{tabular}{|c|l|}
\hline
Observables & Time evolutions \\
\hline
$|A_0(t)|^2$  & $|A_0(0)|^2 \left[e^{-\Gamma_L t} -
e^{-\overline{\Gamma}t}
\sin(\Delta m t)\delta\phi\right]$\\
$|A_{\|}(t)|^2$ &$ |A_{\|}(0)|^2 \left[e^{-\Gamma_L t} -
e^{-\overline{\Gamma}t}\sin(\Delta m t)\delta\phi\right]$\\
$|A_{\perp}(t)|^2$ & $|A_{\perp}(0)|^2 \left[e^{-\Gamma_H t} +
e^{-\overline{\Gamma}t}\sin(\Delta m t)\delta\phi\right]$\\
\hline

Re$(A_0^*(t) A_{\|}(t))$ &  $|A_0(0)||A_{\|}(0)|\cos(\delta_2 -
\delta_1)\left[e^{-\Gamma_L t} - e^{-\overline{\Gamma}t}
\sin(\Delta m t)\delta\phi\right]$\\
Im$(A_{\|}^*(t)A_{\perp}(t))$ & $|A_{\|}(0)||A_{\perp}(0)|\left[
e^{-\overline{\Gamma}t}\sin (\delta_1-\Delta m t)+\frac{1}{2}\left(
e^{-\Gamma_H t}-e^{-\Gamma_L
t}\right)\cos(\delta_1)\delta\phi\right]$\\
Im$(A_0^*(t)A_{\perp}(t))$ & $|A_0(0)||A_{\perp}(0)|\left[
e^{-\overline{\Gamma}t}\sin (\delta_2-\Delta m t)+\frac{1}{2}\left(
e^{-\Gamma_H t}-e^{-\Gamma_L
t}\right)\cos(\delta_2)\delta\phi\right]$\\
\hline
\end{tabular}
\end{center}\caption{Time evolution of the decay $B_s\to J/\psi(\to l^+l^-)
\phi(\to K^+K^-)$ of an initially (i.e.\ at $t=0$) pure $B_s$ meson.}
\label{tab1}
\end{table}

The time-dependent observables in all these
decays provide information about the corresponding values of 
$\Gamma_L$, $\Gamma_H$ and $\Delta m$. If we integrate over
the angles $\varphi$ and $\psi$, the angular distribution in the 
remaining ``transversity'' angle $\theta$ can help in separating
out the CP even and odd components and in determining their 
lifetimes separately \cite{ddlr}.

The width difference $\Delta\Gamma\equiv\Gamma_H-
\Gamma_L$ may be sizeable \cite{deltagamma}. 
Because of this difference, 
the interference effects  between the
CP-even and CP-odd final-state  
configurations give rise to  
a term in the time evolution of the {\it untagged} rate, which is
proportional to 
$\left(e^{-\Gamma_H t}-e^{-\Gamma_L
t}\right) \delta \phi$ \cite{fd}.
Thus, with the angular distribution,
CP violating effects may be observable even without
tagging the flavor of the initial $B$.
This feature may be important, because it provides an alternative to
previous investigations, which have shown how to extract
$\sin\phi_{\mbox{{\scriptsize CKM}}}$ from tagged, time-dependent  
analyses
\cite{cptaggedbs}.

The observables of the angular distributions can be 
determined from experimental data by an angular-moment 
analysis \cite{ddf1, dqstl, also-moments} in which 
the data are weighted by judiciously chosen weighting functions 
in order to arrive {\it directly} at the observables.
In \cite{ddf1}, a method applicable to all kinds of angular 
distributions is  
indicated, where the weighting functions can be determined 
without any {\it a priori} knowledge of the values of the 
coefficients. 
This method is almost as efficient
as the likelihood-fit method for a small number of parameters 
and is expected to stay robust even with
a large number of parameters where the maximum likelihood fit
method may be unreliable \cite{stat}.  

Using appropriate weighting functions
for the angular
distributions of the decay products in the transitions $B_s\to J/
\psi\phi$ and/or $B_s\to D_s^{\ast+}D_s^{\ast-}$, one can extract
$(\Gamma_H,
\Gamma_L, \Delta m)_{B_s}$.
The observables of the angular  
distributions of 
$B_s \to J/\psi \phi, ~ D_s^{*+}D_s^{*-}$
can be related to those of the decays 
$B \to J/\psi K^*,D_s^{*+}\overline{D}^*$ by using the $SU(3)$ 
flavor symmetry, where $B$ stands for $B_d$ or $B^+$. 
Determination of the time-dependent angular distributions
will also be 
useful in testing form factor models \cite{ff}  
and furthermore in determining 
the extent to which factorization or the $SU(3)$ flavor symmetry 
hold in these decays.
The full 
angular distributions for all these transitions are given 
explicitly, and the corresponding weighting functions are 
specified in \cite{ddf1}.

	Whereas $B \to J/\psi K^*$ angular distributions (or the
``gold-plated'' mode  $B \to J/\psi K_S$) give the value of 
$\sin(2 \beta)$, a discrete ambiguity still remains in the 
determination of $\beta$. Using $B_s \to J/\psi \phi$
angular distributions in addition, and using the SU(3)
flavor symmetry (only weakly), this 
ambiguity can be 
resolved \cite{ddf2}.

\section{Model-independent estimations of $B \to PP$ and 
$B \to VP$ amplitudes}

Model-dependent rate calculations of $B \to PP$ and 
$B \to VP$ amplitudes have been made in 
\cite{chau, models, desh-models,  ciu}.
While flavor SU(3) by itself cannot predict the rate of
a process, when combined with the experimental observations
of some decay rates, reliable predictions can be made
about the others. This helps in identifying the decay channels 
which would be expected to display
a significant CP violation.

\subsection{$B \to PP$}
\label{subsec:pp}

	The measured branching ratios (in units of $10^{-5}$)
\cite{cleokpi}
$B(B^0 \to K^+ \pi^-) = 1.4 \pm 0.3 \pm 0.1$ and
$B(B^+ \to K^0 \pi^+) = 1.4 \pm 0.5 \pm 0.2$, when
compared with \cite{cleoketa} 
$B(B^+ \to K^+ \eta') = 7.4^{+0.8}_{-1.3} \pm 0.9$ may imply a 
significant contribution from the flavor
SU(3) singlet component of the $\eta'$ \cite{keta}. 
While one
possibility for this contribution \cite{BRS,kb,int} is an intrinsic 
$c \bar c$
component in the $\eta'$, more conventional mechanisms 
\cite{int,other} (e.g.,
involving gluons) also seem adequate to explain the observed rate.  
Parametrizing this contribution as ${\cal S}$ or ${\cal S'}$
(by using the same 
Feynman diagram basis mentioned in Sec.~\ref{rigid}), 
and adding
the information from the branching ratio of $B \to \pi \pi$,
one can estimate the magnitudes of $\cat, \catp, \cap, \capp,
{\cal S}$ and ${\cal S'}$. Since all the $B \to PP$ decay 
processes are dominated by one or more of these contributions,
the decay rate of any process can thus be estimated.
This has been explicitly done in \cite{keta}. (The 
relative phases between these amplitudes are still unknown, so
only a range for the branching ratio of a decay mode can be given.) 

The potential
for CP-violating rate asymmetries to be exhibited in decays of $B$ mesons to
pairs of charmless mesons has been noted in \cite{Asymms}.
To observe direct CP violation in a decay mode, we 
need at least two significant contributions to that mode, which
have different strong as well as weak phases. 
As a rule of thumb, one must at least
be able to observe the {\it square} of the lesser of the two 
interfering amplitudes
at the $n \sigma$ level in order to observe an asymmetry at this level
\cite{DR}. With the above estimation of the contributions of
different amplitudes, it is seen that this sensitivity threshold 
is passed for
decays of the form $B^+ \to \pi^+ \eta$ and $B^+ \to \pi^+ \eta'$ when
branching ratios of order $10^{-6}$ become detectable in experiments 
sensitive
to both charged and neutral final-state particles. These two modes
thus emerge as promising ones for observing direct CP
violation \cite{keta}.

\subsection{$B \to VP$}

The observation of the decays $B^+ \to \omega \pi^+$ and $\omega K^+$ at
branching ratio levels of about $10^{-5}$ by the CLEO Collaboration
\cite{CLEOVP} can be used, with the help of flavor SU(3), to anticipate the
observability of other charmless $B \to VP$ decays in the near future  
\cite{vp}.

Now we need to have twice the number of amplitudes for our basis.
The amplitudes will depend upon whether the spectator quark 
(the quark other than
the $b$ in the decaying $B$ meson) ends up in the final state
scalar or vector meson. The amplitudes contributing to a significant
extent will then be denoted by $t_P, t_V, c_P, c_V, 
p_P, p_V, s_P, s_V$ (and their primed counterparts for
$|\Delta S| =1$ decays)
where the subscript denotes where the spectator quark goes.

One can then study the hierachy of these amplitudes, on similar
lines to the hierarchy in $B \to PP$ mentioned in
Sec.~\ref{rigid}. 
The fact that the
$B^+ \to \omega \pi^+$ and $B^+ \to \omega K^+$ branching ratios are comparable
to one another and each of order $10^{-5}$ indicates that the dominant
contribution to $\omega \pi^+$ is most likely $t_V$, while the dominant
contribution to $\omega K^+$ is most likely $p'_V$.  
An appreciable value for the amplitude $p'_V$, somewhat of a surprise on the
basis of conventional models 
\cite{chau, desh-models, ciu}, implies that the decays $B \to \rho K$ should
be observable at branching ratio levels in excess of $10^{-5}$.  The smallness
of the ratio ${\cal B}(B^+ \to \phi K^+)/{\cal B}(B^+ \to \omega K^+)$
indicates that $|p'_P| < |p'_V|$.  The amplitude $p'_P$ should dominate not
only $B \to \phi K$ but also $B \to K^* \pi$ decays.  Evidence for any of these
would then tell us the magnitude of $p'_P$.  The relative phase of $p'_P$ and
$p'_V$ is probed by $B \to K^* (\eta,\eta')$ decays.

The amplitude $s'_V$, coupling to the flavor SU(3) singlet component of
the $\eta$  and $\eta'$, can be as large as or even larger than $p'_V$. 
Several tests can be suggested
for non-zero singlet amplitudes, including a number of triangle and rate
relations. A program for determining the magnitude and
phase of $s'_V$ has also been outlined in \cite{vp}.

Once the dominant $t_V$, $p'_V$, and $s'_V$ 
amplitudes have been determined, one
can use flavor SU(3) to predict the amplitudes $t'_V$, $p_V$, and $s_V$.  
It is
then a simple matter (along the lines indicated in 
Sec. \ref{subsec:pp}) 
to determine which processes have
the potential for exhibiting noticeable interferences between two or more
amplitudes, and thereby displaying CP-violating
asymmetries.

\section{SU(3) Breaking}

Flavor SU(3) is not an exact symmetry and while making 
predictions on its basis, the errors due to the SU(3) 
breaking effects need to be estimated and the corrections
need to be taken into account. There is no
theoretically clean way of calculating these corrections,
however a parametrization has been proposed
in \cite{ghlrsu3} in the basis of Feymnan diagrams, where
the implications of the SU(3) breaking effects for the 
extraction of weak phases are also examined.

Within the framework of generalized factorization, the SU(3)
breaking can be looked upon as arising
 from the following sources :
\begin{itemize}
\item different decay constants of the final state particles, 
	$f_P$ and $f_V$
\item different form factors 
\item different masses of the final state particles.
\end{itemize}
A study of the SU(3) breaking corrections using generalized 
factorization is currently in progress \cite{addd}.

\section*{Acknowledgements}

We thank Mehta Research Institute, Allahabad for their hospitality
during the Workshop on $B$ and Neutrino Physics. The topics
discussed in this talk are based on the work carried out with
I. Dunietz, R. Fleischer, M. Gronau, H. Lipkin, J. L. Rosner,
and S. Sen. We would like to thank them for the enjoyable
collaborations.


\begin{thebibliography}{99}


\bibitem{ckm}N. Cabibbo, Phys. Rev. Lett. {\bf 10}, 531 (1963); M.
Kobayashi
and T. Maskawa, Prog. Theor. Phys. {\bf 49}, 652 (1973); {\em B
Decays}, edited
by S. Stone, World Scientific, 1994.

\bibitem{zepp} D. Zeppenfeld, \zpc{8}{77}{1981}; 

\bibitem{savagesu3} M. Savage and M. Wise, \prd{39}{3346}{1989};
\ibj{40}{3127(E)}{1989}; 

\bibitem{chau} L. L. Chau {\it et al.}, 
\prd{43}{2176}{1991}.

\bibitem{glr} M. Gronau and D. London, Phys. Rev. Lett. {\bf 65},
3381 (1990); M. Gronau, J. L. Rosner, and D. London, Phys. Rev.
Lett. {\bf 73}, 21 (1994).
 
\bibitem{ghlrd50}M. Gronau, O. F. Hernandez, D. London, and J. L.
Rosner, Phys. Rev. D {\bf 50}, 4529 (1994).
 
\bibitem{desh-he}N. G. Deshpande and X.-G. He, Phys. Rev. Lett.
{\bf 75}, 3064 (1995).

\bibitem{triangles}A. J. Buras and R. Fleischer, Phys. Lett.
{\bf B360}, 138 (1995);
M. Gronau and J. L. Rosner, \prd{53}{2516}{1996}.

\bibitem{hierarchy} M. Gronau, O. Hernandez, D. London, and J. L.
Rosner, Phys. Rev. D {\bf 52}, 6374 (1995).

\bibitem{art}  A. S. Dighe, \prd{54}{2067}{1996}.

\bibitem{etaetap}  A. S. Dighe, M. Gronau, and J. L. Rosner, 
\plb{367}{357}{1996}; \ibj{377}{325(E)}{1996}.

\bibitem{charmedP} A. J. Buras and R. Fleischer, Phys. Lett. {\bf
B341},
379 (1995).

\bibitem{Browder} T. Browder, K. Honscheid and S. Playfer, in \stone, p.~158.

\bibitem{BS} D. Bortoletto and S. Stone, \prl{65}{2951}{1990};
J. L. Rosner, Phys. Rev. {\bf D42}, 3732 (1990).

\bibitem{Penphase} M. Bander, D. Silverman and A. Soni, \prl{43}{242}{1979};
G. Eilam, M. Gronau and J. L. Rosner, \prd{39}{819}{1989}; L. Wolfenstein,
\prd{43}{151}{1991} J.-M. G\'erard and W.-S. Hou, \prd{43}{2909}{1991};
H. Simma, G. Eilam and D. Wyler, \np{352}{367}{1991}.

\bibitem{gr} M. Gronau and J. L. Rosner, Phys. Rev. Lett. {\bf 76},
1200 (1996).

\bibitem{dgr} A. S. Dighe, M.
Gronau, and J. L. Rosner, \prd{54}{3309}{1996}.

\bibitem{discrete} A. S. Dighe and J. L. Rosner,
\prd{54}{4677}{1996}.

\bibitem{london} C.S. Kim, D. London and T. Yoshikawa, 
Phys. Rev. D {\bf 57}, 4010 (1998).


\bibitem{angular} G. Valencia, Phys.\ Rev.\ {\bf D39}, 3339 (1989);
G. Kramer and W.F. Palmer,
Phys.\ Rev.\ {\bf D45}, 193 (1992),
Phys.\ Lett.\ {\bf B279}, 181(1992),
Phys.\ Rev.\ {\bf D46} 2969 and 3197 (1992);
G. Kramer, W.F. Palmer and H. Simma, Nucl.\ Phys.\ {\bf B428}
77 (1994);
G. Kramer, T. Mannel and W.F. Palmer, Z.\ Phys.\ {\bf C55}
497 (1992).

\bibitem{ddlr}
 A. S. Dighe, I. Dunietz, H. J. Lipkin, and J. L. Rosner,
\plb{369}{144}{1996}.

\bibitem{ddf1} A. S. Dighe, I. Dunietz and R. Fleischer, 
in preparation. (Since this talk, issued as hep-ph/9804253, 
to be published in Eur. Phys. J. C.) 

\bibitem{dphi}R. Aleksan, B. Kayser and D. London, Phys.\ Rev.\
Lett.\ {\bf 73} (1994) 18.

\bibitem{deltagamma}
J.S. Hagelin, Nucl.\ Phys.\ {\bf B193}, 123 (1981);
E. Franco, M. Lusignoli and A. Pugliese, {\it ibid.}~{\bf B194},
403 (1982); L.L Chau, W.-Y. Keung and M. D. Tran, Phys.\ Rev.\
{\bf D27}, 2145 (1983); L.L Chau, Phys.\ Rep.\ {\bf 95}, 1 (1983); 
A. J. Buras, W. Slominski and H.~Steger, Nucl.\ Phys.\ {\bf B245},
369 (1984); M.B. Voloshin, N.G. Uraltsev, V.A. Khoze and M.A.
Shifman, Yad.\ Fiz.\ {\bf 46}, 181 (1987) [Sov.\ J. Nucl.\ Phys.\
{\bf 46}, 112 (1987)]; A.~Datta, E.A. Paschos and U. T\"urke,
Phys.\ Lett.\ {\bf B196}, 382 (1987); A. Datta, E.A..~Paschos and
Y.L. Wu, Nucl.\ Phys.\ {\bf B311}, 35 (1988); M. Lusignoli, Z.
Phys.\ {\bf C41}, 645 (1989); R. Aleksan, A. Le Yaouanc, L. Oliver,
O. P\`ene and Y.-C. Raynal, Phys.\ Lett.\ {\bf B316}, 567 (1993);
I. Bigi, B. Blok, M. Shifman, N. Uraltsev and A.~Vainshtein,
in {\it B Decays}, ed.\ S. Stone, 2nd edition (World
Scientific, Singapore, 1994), p.\ 132 and references
therein; M. Beneke, G. Buchalla and I. Dunietz, Phys.\ Rev.\ {\bf D54},
4419 (1996).

\bibitem{fd}R. Fleischer and I. Dunietz, Phys.\ Rev.\ {\bf D55},
259 (1997); R. Fleischer and I. Dunietz, Phys.\ Lett.\ {\bf B387},
361 (1996).


\bibitem{cptaggedbs}
D. Du, I. Dunietz, and Dan-di Wu, Phys.\ Rev.\ {\bf D34}, 3414 (1986);
I. Dunietz and J.L.~Rosner, Phys.\ Rev.\ {\bf D34}, 1404 (1986);
Ya.I. Azimov, N.G. Uraltsev and V.A.~Khoze, JETP Lett. {\bf 43}, 409
(1986);
I. Dunietz, Ann. Phys. {\bf 184}, 350 (1988);
I. Dunietz, in Proceedings of the Workshop on $B$ Physics at Hadron
Accelerators, Snowmass, Colorado, June 21 - July 2, 1993, p. 83, eds.\
P.~McBride and C.~Shekhar~Mishra, Fermilab-CONF-93/267.


\bibitem{dqstl} I. Dunietz, H. R. Quinn, A. Snyder, W. Toki and
H. J. Lipkin, Phys.\ Rev.\ {\bf D43},
2193 (1991).

\bibitem{also-moments}
P. Bialas et al., Z.\ Phys.\ {\bf C57}, 115 (1993);
J. Hrivnac, R. Lednicky and M. Smizanska, J. Phys.~{\bf G21}, 629 (1995);
Q. Shen et al., Phys.\ Rev.\ {\bf D52}, 2825 (1995).

\bibitem{stat} A. Dighe and S. Sen, in preparation. 
(Since this talk, issued as ICTP preprint IC/98/177,
hep-ph/9810381.)

\bibitem{ff} M. Bauer, B. Stech and M. Wirbel, Z. Phys.\ {\bf C29},
637 (1985) and Z. Phys.\ {\bf C34}, 103 (1987);
J.M. Soares, Phys.\ Rev.\ {\bf D53}, 241 (1996);
H.-Y. Cheng, Z. Phys.\ {\bf C69}, 647 (1996).

\bibitem{ddf2} A. S. Dighe, I. Dunietz and R. Fleischer, 
in preparation. (Since this talk, published in
Phys. Lett. {\bf B433}, 147 (1998).)


\bibitem{models} M. Bauer, B. Stech, and M. Wirbel, \zpc{34}{103}{1987};
A. DeAndrea, N. Di Bartolomeo, R. Gatto, and G. Nardulli, \plb
{318}{549}{1993};
A. DeAndrea, N. Di Bartolomeo, R. Gatto, F. Feruglio, and G.
Nardulli, \plb{320}{170}{1994};
G. Kramer and W. F. Palmer, \prd{52}{6411}{1995};
\zpc{66}{429}{1995};
R. Aleksan \ite, \plb{356}{95}{1995};
H.-Y. Cheng and B. Tseng, IP-ASTP-03-97/NTU-TH-97-08,
hep-ph/9707316; IP-ASTP-04-97/NTU-TH-97-09, hep-ph/9708211;

\bibitem{desh-models} 
N. G. Deshpande and J. Trampeti\'{c}, \prd{41}{895}{1990}; see
also N. G. Deshpande in \stone, p.~587.

\bibitem{ciu} M. Ciuchini, E. Franco, G. Martinelli, and L. Silvestrini,
CERN report CERN-TH/97-30, hep-ph/9703353; M. Ciuchini, R. Contino, E. Franco,
G. Martinelli, and L. Silvestrini, CERN report CERN-TH/97-188, hep-ph/9708222.
This last work contains an extensive list of further references.

\bibitem{cleokpi} CLEO Collaboration, CLEO CONF 98-20, ICHEP98 858.

\bibitem{cleoketa}CLEO Collaboration, CLEO CONF 98-09, ICHEP98 860.

\bibitem{keta} A. S. Dighe, M. Gronau
      and J. L. Rosner, Phys. Rev. Lett. {\bf 79}, 4333 (1997). 

\bibitem{BRS} S. Barshay, D. Rein, and L. M. Sehgal, \plb{259}{475}{1991}.

\bibitem{kb} K. Berkelman, CLEO note CBX 96-79 and supplement, unpublished.

\bibitem{int} I. Halperin and A. Zhitnitsky, hep-ph/9704412 
and hep-ph/9705251;
F. Yuan and K.-T. Chao, \prd{56}{R2495}{1997}; A. Ali and C. Greub, DESY
97-126, hep-ph/9707251.


\bibitem{other} D. Atwood and A. Soni, \plb{405}{150}{1997}; hep-ph/9706512;
W.-S. Hou and B. Tseng, hep-ph/9705304; H.-Y. Cheng and B. Tseng,
IP-ASTP-03-97/NTU-TH-97-08, hep-ph/9707316; A. Datta, X.-G. He, and S. Pakvasa,
hep-ph/9707259; A. L. Kagan and A. A. Petrov, UCHEP-27/UMHEP-443,
hep-ph/9707354; H. Fritzsch, CERN-TH/97-200, hep-ph/9708348.


\bibitem{Asymms} M. Bander, D. Silverman, and A. Soni, \prl{43}{242}{1979}; J.
M. G\'erard and W. S. Hou, \prl{62}{855}{1989}; \prd{43}{2909}{1991};
G. Kramer, W. F. Palmer, and H. Simma, \np{B428}{77}{1994}; \zpc{66}{429}
{1995}.

\bibitem{DR} I. Dunietz and J. L. Rosner, \prd{34}{1404}{1986}.


\bibitem{CLEOVP} CLEO Collaboration, CLEO CONF 97-23, EPS 335

\bibitem{vp} A. S. Dighe,  M. Gronau and J. L. Rosner, Phys. Rev. D
{\bf 57}, 1783 (1998).


\bibitem{ghlrsu3} M. Gronau, O. Hern\'andez, D. London, and J. L. Rosner,
\prd{52}{6356}{1995}.

\bibitem{addd} R. Aleksan, N. Deshpande, A. Dighe and B. Dutta,
in preparation.

\end{thebibliography}
\end{document}